\documentclass[aps,twocolumn,prl,
groupedaddress,nofootinbib,nobalancelastpage,nobibnotes,longbibliography]{revtex4-1}
\pdfoutput=1
\usepackage{amsmath,amsfonts,amssymb,mathrsfs,graphicx,color}

\newcommand{\ie}{{\it i.e.}}

\newcommand{\eg}{{\it e.g.}}

\newcommand{\cf}{{\it cf.}}

\newcommand{\eq}{Eq.}

\newcommand{\fig}{Fig.}

\newcommand{\Ref}{Ref.}
\newcommand{\Refs}{Refs.}
\newcommand{\Sec}{Section}

\newcommand{\equ}[1]{\eq~(\ref{equ:#1})}
\newcommand{\figu}[1]{\fig~\ref{fig:#1}}

\newcommand{\bi}{\begin{itemize}}
\newcommand{\ei}{\end{itemize}}

\begin{document}

\title{Neutrino and Cosmic-Ray Emission from Multiple Internal Shocks \\ In Gamma-Ray Bursts}

\author{Mauricio Bustamante}
\email{bustamanteramirez.1@osu.edu}
\affiliation{Center for Cosmology and AstroParticle Physics (CCAPP), The Ohio State University, 191 W.~Woodruff Ave.,
        Columbus, OH 43210, USA}
\affiliation{DESY, Platanenallee 6, D-15738 Zeuthen, Germany}
\affiliation{Institut f{\"u}r Theoretische Physik und
  Astrophysik, Universit{\"a}t W{\"u}rzburg, Am Hubland, D-97074 W{\"u}rzburg, Germany}

\author{Philipp Baerwald}
\affiliation{Department of Astronomy and Astrophysics; Department of Physics; Center for Particle and Gravitational Astrophysics; Institute for Gravitation and the Cosmos; Pennsylvania State University, 525 Davey Lab, University Park, PA 16802, USA}

\author{Kohta Murase}
\affiliation{Department of Astronomy and Astrophysics; Department of Physics; Center for Particle and Gravitational Astrophysics; Institute for Gravitation and the Cosmos; Pennsylvania State University, 525 Davey Lab, University Park, PA 16802, USA}
\affiliation{Hubble Fellow -- Institute for Advanced Study, Princeton, New Jersey 08540, USA}

\author{Walter Winter}
\affiliation{DESY, Platanenallee 6, D-15738 Zeuthen, Germany}

\date{\today}

\begin{abstract}

Gamma-ray bursts are short-lived, luminous explosions at cosmological distances, thought to originate from relativistic jets launched at the deaths of massive stars. They are among the prime candidates to produce the observed cosmic rays at the highest energies. Recent neutrino data have, however, started to constrain this possibility in the simplest models with only one emission zone.
In the classical theory of gamma-ray bursts, it is expected that particles are accelerated at mildly relativistic shocks generated by the collisions of material ejected from a central engine.  
We consider neutrino and cosmic-ray emission from multiple emission regions since these internal collisions must occur at very different radii, from below the photosphere all the way out to the circumburst medium, as a consequence of the efficient dissipation of kinetic energy.  
We demonstrate that the different messengers originate from different collision radii, which means that multi-messenger observations open windows for revealing the evolving GRB outflows.

\end{abstract}


\maketitle


Gamma-ray bursts (GRBs) are violent outbreaks of energy distributed over cosmological distances. Most of the energy is detected as gamma rays during the so-called prompt phase, lasting from a few seconds to several hundred seconds (see \Refs~\cite{Piran:2004ba,Meszaros:2006rc,Zhang:2007nka} for reviews). The common view is that relativistic jets are ejected from a central engine, triggered by a collapsing star or a neutron star merger, in the direction of the observer.  The inhomogeneity in the jets naturally leads to internal shocks, at which charged particles can be accelerated.  
In the classical GRB scenario, the observed gamma-ray emission is attributed to synchrotron radiation from nonthermal electrons.
It is natural to expect that protons are accelerated as well, and GRBs have also been considered as a possible candidate class for the origin of the ultra-high-energy cosmic rays (UHECRs)~\cite{Milgrom:1995um,Waxman:1995vg,Vietri:1995hs}.  Whereas the charged cosmic rays cannot be traced back to their origin because of their deflection on magnetic fields during propagation, neutrinos from GRBs, which would be generated via proton-gas or proton-radiation interactions, point back to the sources and could provide crucial clues to the UHECR mystery~\cite{Waxman:1997ti}.  

Neutrinos up to PeV energies from presumably extragalactic sources have now been detected in the IceCube neutrino telescope~\cite{Aartsen:2013jdh}. While even the signal shape seems compatible with a GRB origin~\cite{Murase:2005hy,Murase:2008sp,Baerwald:2010fk,Zhang:2012qy}, stacking searches for prompt GRB neutrinos using the timing and directional information coming from gamma-ray observations have been so far unsuccessful~\cite{Aartsen:2013bka,Abbasi:2012zw}.  Because some of the early analytical predictions of the GRB neutrino fluxes~\cite{Guetta:2003wi,Abbasi:2012zw} have shortcomings which are independent of astrophysical uncertainty (although these do not exist in some numerical works such as \Ref~\cite{Murase:2005hy}), the model used by IceCube in \Ref~\cite{Abbasi:2012zw} has been revised by about one order of magnitude~\cite{Hummer:2011ms,Li:2011ah,He:2012tq}. The current data are even pushing into the expected regime of the latest predictions, enabling us to address whether GRBs can be the sources of the UHECRs, and what the neutrinos can tell us about that.  

In most of the earlier discussions, a simple one-zone model is assumed: this approach considers one representative collision between two relativistic plasma blobs representing the inhomogeneity in the jet, calculates the emission from this collision, and scales the result for the whole burst by assuming many such {\it identical} collisions within the jet. In this simple model, the GRB parameters are fixed during its duration.  In particular, the internal shock radius $R_{\text{C}}$, where the representative collision occurs and which is crucial for neutrino and UHECR production \cite{Murase:2005hy}, is often estimated from geometric arguments~\cite{Halzen:2002pg}. Taking the blobs as spherical shells, and using the representative value of the Lorentz factor $\Gamma \equiv \left(1-\left(v/c\right)^2\right)^{-1/2}$ of the plasma blobs, with $v$ the average velocity of the blobs, the variability timescale $t_{\text{v}}$, and the burst redshift $z$, the collision radius can be estimated as 
\begin{equation}
R_{\text{C}} \simeq 2 \, \Gamma^2 \, c \, t_{\text{v}} /(1+z) \, . \label{equ:rc}
\end{equation}
The variability timescale can be obtained by inspection of the pulse rising time of the burst's light curve; the Lorentz factor can only be estimated, using various approaches~\cite{Piran:1999kx,2010MNRAS.402.1854Z}; and the redshift can be estimated via the observation of the host galaxy of the GRB.
In the internal shock model, using the typical variability timescale (which is about three times shorter than the pulse width of $\sim 1 \, \mathrm{s}$~\cite{Nakar:2001iz,Bhat:2013fsa}), $R_{\text{C}} \sim 10^8$--$10^{10.5} \, \mathrm{km}$ is expected~\cite{Nakar:2002gd} and neutrino predictions correspondingly vary~\cite{Murase:2005hy,Murase:2008mr,He:2012tq}. 
Specifically, in dissipative photospheric scenarios~\cite{Murase:2008sp,Wang:2008zm,Murase:2013hh,Bartos:2013hf,Gao:2012ay}, internal shocks may occur under or around the radius known as the photosphere, at which the Thomson optical depth for $e\gamma$ scattering~\cite{Rees:2004gt} is unity. Gamma rays can directly escape above the photosphere, where the optical depth is low. Even beyond it, high-energy gamma rays are attenuated by $\gamma\gamma$ interactions. Since the photospheric radius $R_{\rm ph} \sim 10^{8}$--$10^{8.5} \, \mathrm{km}$ is small, neutrino production is expected to be highly efficient around the photosphere.  

UHECR production is also sensitive to $R_{\text{C}}$; UHECR escape also depends on GRB parameters~\cite{Baerwald:2013pu}. Although it is often assumed that UHECRs can escape after the dynamical timescale (\ie, the shock crosses the shell), this is not the case if magnetic fields do not decay. Especially strong constraints on the UHECR-neutrino connection can be obtained if cosmic rays escape only as neutrons which are produced in the same interactions as the neutrinos (``neutron-escape model'')~\cite{Rachen:1998fd,Mannheim:1998wp}. While this specific model is essentially excluded~\cite{Baerwald:2014zga,Ahlers:2011jj}, a hard flux of protons leaking from the sources (hereafter called ``direct escape'' around the maximum energy and/or ``diffusion escape'' at lower energies) can dominate the UHECR emission, which is largely allowed by neutrino observations~\cite{Baerwald:2013pu,Baerwald:2014zga}. As demonstrated in \Ref~\cite{Baerwald:2013pu}, the dominant UHECR escape mechanism is in fact a function of the shell parameters.  

Since the one-zone model is not realistic in the internal shock picture, $R_{\text{C}}$ and $\Gamma$ should evolve even within one GRB.  The $R_{\text{C}}$-dependence of neutrino production efficiency has been discussed in the internal shock model~\cite{Guetta:2001cd,Murase:2008mr,He:2012tq,Zhang:2012qy}, but its integrated effects have not been studied in detail.  Hence, it is conceivable that not all collisions occur at the same radius, which has significant consequences for the neutrino and cosmic-ray production, as we show in this work. For example, different UHECR escape mechanisms will dominate in different phases of the evolving GRB. 

We demonstrate that the different messengers originate from different collision radii. Even in the internal shock model, the neutrino production can be dominated by emission from around the photosphere, \ie, the radius where the ejecta become transparent to gamma-ray emission.  Possible subphotospheric contributions enhance the detectability.  We predict a minimal neutrino flux per flavor at the level of $E^2 J \sim 10^{-11} \, \mathrm{GeV \, cm^{-2} \, sr^{-1} \, s^{-1}}$ for the contribution from beyond the photosphere, with a spectral shape similar to the original theoretical prediction.  However, in striking contrast to earlier approaches, this prediction turns out to hardly depend on model parameters such as the Lorentz boost, the baryonic loading, or the variability timescale.

\section{Results}

\subsection{Dynamical burst model}

\begin{figure}[t!]
\begin{center}
\includegraphics[width=\columnwidth]{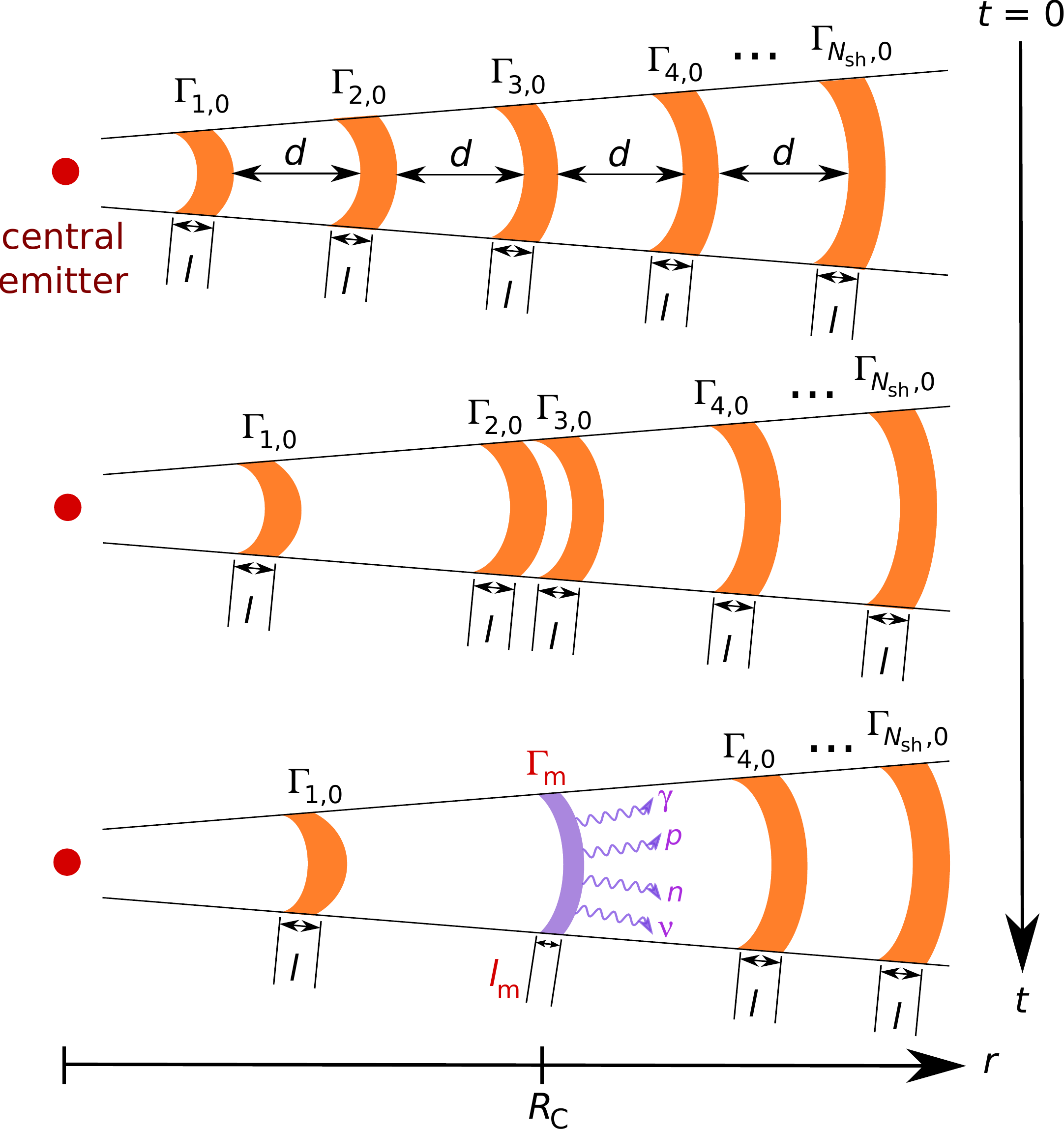}
\end{center}
\caption{\label{fig:Figure-1-Bustamante}{\bf Illustration of the internal-collision model of gamma-ray bursts used in this study.} A set of $N_{\mathrm{sh}}$ shells with equal energies, widths and separations $l = d = c {\delta t}_{\rm eng}$ are emitted from a central engine, where ${\delta t}_{\rm eng}$ is the uptime of the central emitter. The shells have a spread in the bulk Lorentz factor but initially equal bulk kinetic energies. The shells propagate, collide, and merge (marked by the shell colored purple) as soon as they meet other shells (multiple collisions are allowed), whereupon their masses, widths and speeds change. The energy dissipated in the collision is assumed to be radiated away immediately.}
\end{figure}

In order to demonstrate neutrino and cosmic-ray emission from various $R_{\text{C}}$, we follow the internal collision model of \Ref~\cite{Kobayashi:1997jk}; see \figu{Figure-1-Bustamante} for illustration.  
We set out a number $N_{\mathrm{sh}}$ of shells from a central engine with equal initial kinetic energies but a spread in the bulk Lorentz factor, around $\Gamma_0$, of
\begin{equation}
\ln \left( \frac{\Gamma_{k,0}-1}{\Gamma_0-1} \right) = A_\Gamma x \; ,
\end{equation}
where $\Gamma_{k,0}$ is the initial Lorentz factor of the $k$-th shell and $x$ follows a Gaussian distribution $P(x) dx = e^{-x^2/2}/\sqrt{2 \pi} dx$. Note that a large dispersion $A_\Gamma \gg 0.1$ is required in order to achieve high efficiencies~\cite{Kobayashi:2001iq}, as we have explicitly tested, since the energy dissipation is proportional to the difference between the Lorentz factors of the colliding shells.  
The shells are assumed to be emitted with an uptime of the emitter $\delta t_{\mathrm{eng}}$, followed by an equally-long downtime, which is an input of the simulation. The variability timescale $t_{\text{v}}$ will be obtained after running the simulation from the light curve as an output, with a value which is typically similar to  $\delta t_{\mathrm{eng}}$. For simplicity, we have assumed constant uptime and downtime, but \Ref~\cite{Nakar:2002gd} explored a scenario where ${\delta t}_{\rm eng}$ is different for each emitted shell and follows a log-normal distribution; post-simulation, it is possible also in this case to infer a variability timescale for the whole burst. While the shells evolve, their widths, masses, and speeds (\ie, their Lorentz factors) are assumed to be constant, and their mass density decreases $\propto r^{-2}$, with $r$ the radial distance to the emitter. Because of the different speeds of the shells, a shell will collide with another and merge into a new one; see \figu{Figure-1-Bustamante}. During the burst evolution, shells may collide several times. We assume that after a collision the new shell immediately cools by prompt radiation of the internal energy into gamma rays, cosmic rays, and neutrinos.  Derivations of the properties of the newly formed shell are given in \Refs~\cite{Kobayashi:1997jk,Aoi:2009ty} and are maintained in the simulations presented here. 
Our results match the analytical predictions for the dissipation of modest-amplitude fluctuations from \Refs~\cite{Beloborodov:2000nn,Li:2008ub}.   
Note that we simplify the evolution of the internal shocks in several points, although our approach is enough for the purpose of this work. First, since we focus on the classical internal shock scenario where optically-thin synchrotron emission is the most relevant mechanism, we assume situations where most of the dissipation occurs in the optically-thin regime. If significant dissipation occurs in the optically-thick regime, the internal energy scales adiabatically $\propto r^{-2/3}$, which is spent to accelerate the outflow. Second, since we do not consider cases where only a fraction of the internal energy made available after a collision is released as radiation~\cite{Kobayashi:2001iq}, this means the efficiency issue of the internal shock model may remain unresolved~\cite{Kumar:1999cv}. Third, observed light curves from real GRBs may have slow variability components as well as fast variability components~\cite{Zhang:2010jt}, which are not easily explained by a discrete number of shells from a continuous emitter, whereas continuous outflow models give better agreement~\cite{Daigne:1998xc,Daigne:2003tp,Bosnjak:2008bd}. 

In this study, we choose for our base model the parameter values $\Gamma_0=500$, $N_{\mathrm{sh}}=1000$, ${\delta t}_{\rm eng}=0.01 \, \mathrm{s}$, and $A_\Gamma=1$, as well as a perfect acceleration efficiency of $\eta=1$ (defined by $t'^{-1}_{\mathrm{acc}} \equiv \eta \, c^2 \, e \, B'/E_p'$, with $E_p'$ the proton energy; see \Ref~\cite{Baerwald:2013pu}). The simulation yields 990 collisions, $t_{\text{v}} \simeq 0.06 \, \mathrm{s}$ from the average obtained rise time of the light curve pulses (see \Ref~\cite{Kobayashi:1997jk}), a burst duration $T \simeq N_{\mathrm{coll}} t_{\text{v}} \approx 59 \, \mathrm{s}$, and an average $\langle\Gamma\rangle \approx 370$ (average Lorentz factor of the merged shells, corresponding to the observable $\Gamma$), \ie, the GRB is sufficiently close to conventional assumptions in neutrino production models. Our study focuses on long-duration GRBs, which typically last tens of seconds, and our chosen parameter sets indeed yield burst durations of that order. We normalize the total isotropic photon energy of all collisions in the source frame to $E_{\mathrm{iso}} = 10^{53} \, \mathrm{erg}$, consistent with GRB observations. Note that the fraction of photon energy dissipated in subphotospheric collisions is only about 9\%, which means that a renormalization of the gamma-ray energy output to only collisions above the photosphere would hardly 
affects our result. For the cosmic-ray and neutrino production, we follow \Refs~\cite{Baerwald:2011ee,Baerwald:2013pu} to compute the spectra for each collision individually, choosing equal energies in electrons (\ie, photons) and magnetic field, and a baryonic loading of ten (\ie, ten times more dissipated energy in protons than in photons). Neutrinos are produced in $p\gamma$ interactions.
The target photon spectrum is assumed to be a broken power law with spectral indices $\alpha_\gamma = 1$ and $\beta_\gamma = 2$, respectively, with a fixed break energy of $\epsilon_{\gamma,\text{break}}^\prime = 1 \, \mathrm{keV}$ in the merged-shell rest frame (primed quantities are in the merged-shell rest frame). That is, it is implied that the target photon spectrum corresponds to conventional GRB observations regardless of the underlying radiation processes leading to this spectral shape.

\begin{figure}[t]
\begin{center}
\includegraphics[width=1.0\columnwidth]{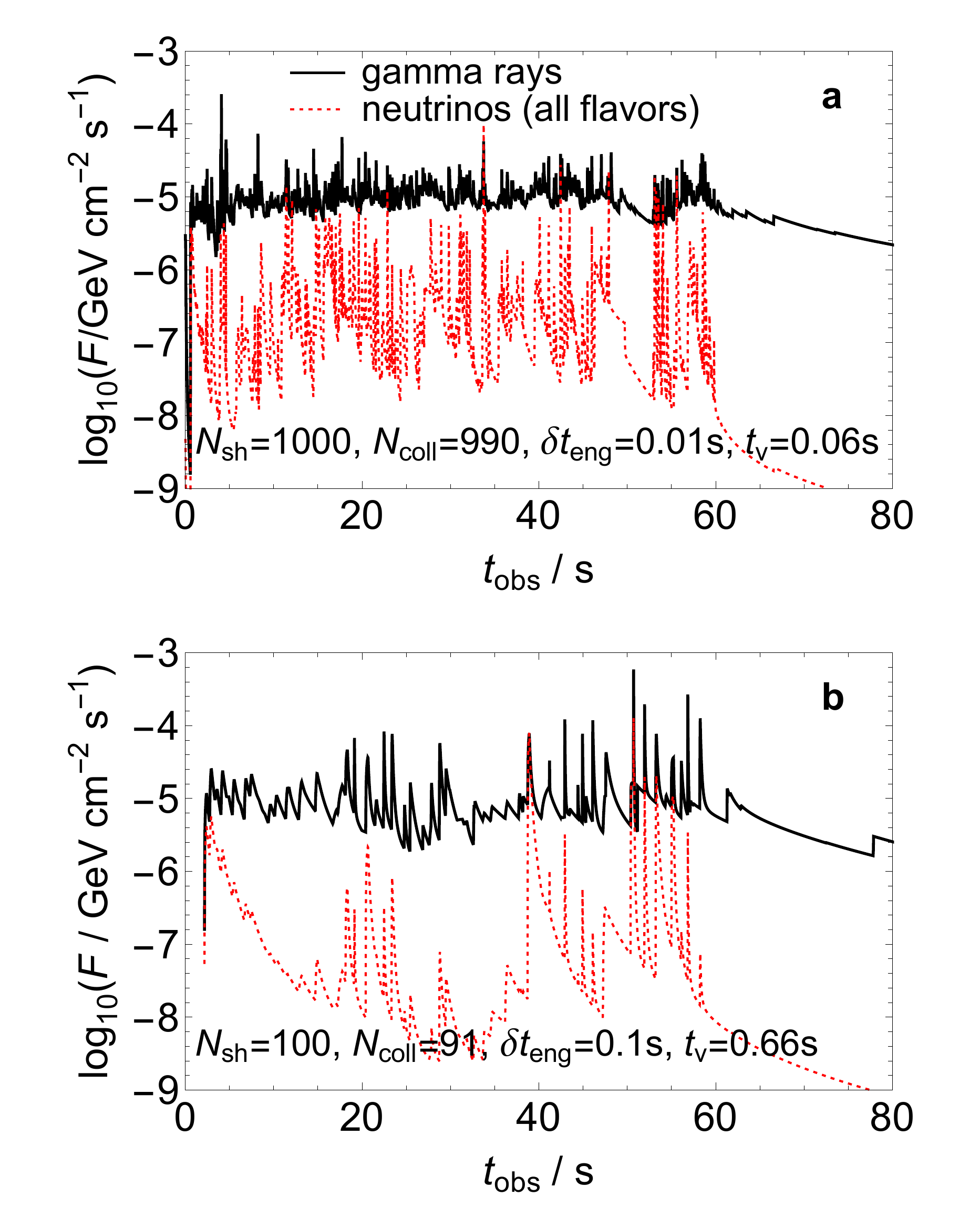}
\end{center}
\caption{\label{fig:Figure-2-Bustamante}
{\bf Two simulated light curves.} The curves for the energy flux of gamma rays (solid, black) and neutrinos (dotted, red) are built from the collisions of shells output by an engine emitting shells with equal kinetic energies, with $t_\text{obs}$ the time in the observer's frame. The light curves in set (a) (set (b)) correspond to a simulation with $N_\text{sh} = 1000$ ($100$), $N_\text{coll} = 990$ ($91$), ${\delta t}_{\rm eng} = 0.01$ s ($0.1$ s), and $t_{\text{v}} = 0.06$ s ($0.66$ s). A redshift of $z=2$ was assumed to produce these light curves.}
\end{figure}

\begin{figure*}[t]
\begin{center}
\begin{tabular}{cccc}
\includegraphics[width=\textwidth]{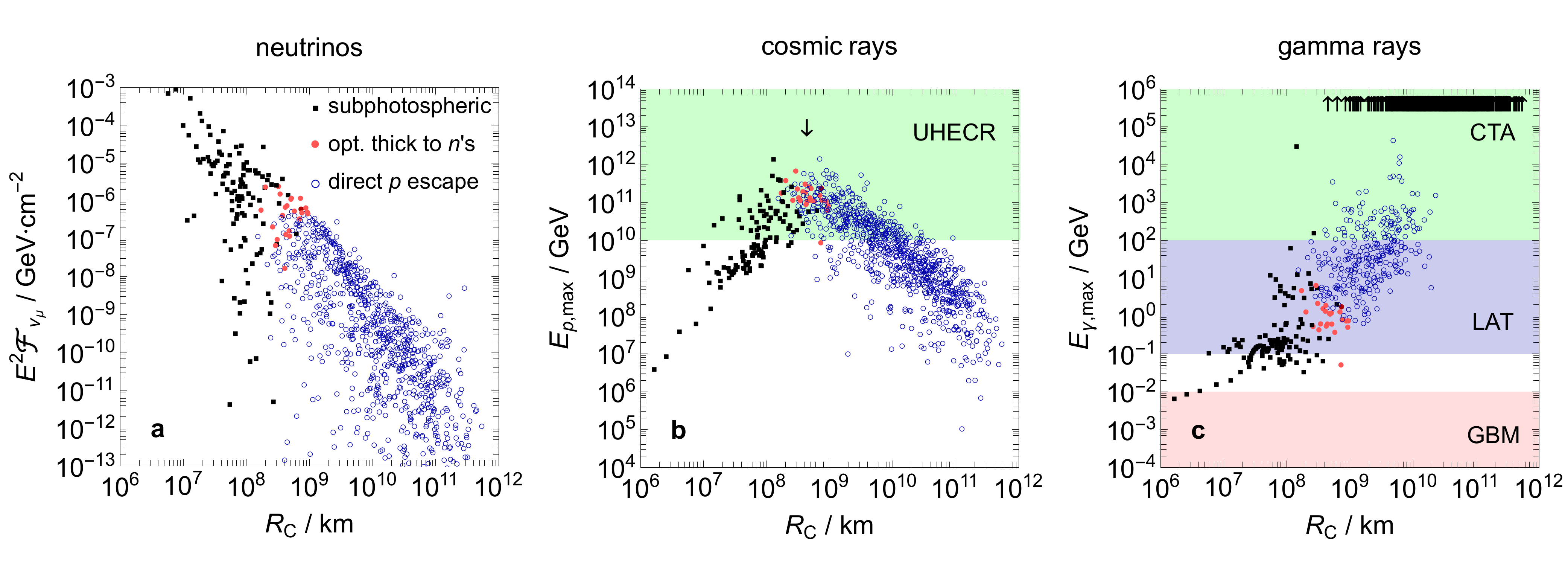}
\end{tabular}
\end{center}
\caption{\label{fig:Figure-3-Bustamante}{\bf Neutrino fluence, maximum proton and gamma-ray energy for each internal collision.} (a) Muon-neutrino fluence ($\nu_\mu+\bar{\nu}_\mu$, in the observer's frame), (b) maximal proton energy (in the source frame, for ideal ($\eta=1$) acceleration), and (c) maximal allowed gamma-ray energy (in the source frame, where $\tau_{\gamma \gamma}(E_{\gamma,\mathrm{max}})=1$) as a function of the collision radius. Each dot represents one collision: red filled dots represent collisions where cosmic rays mainly escape as neutrons (optical thickness to $p\gamma$ interactions larger than unity), blue empty circles represent collisions where cosmic-ray leakage dominates over the neutron-escape model, and black squares denote subphotospheric collisions or collisions where this picture cannot be maintained (\ie, where the Thomson optical depth is large). In panel (b), the ultra-high energy range for cosmic rays, above $10^{10}$ GeV, is shown as a green band; the downward-pointing arrow there marks the approximate energy above which adiabatic energy losses dominate. In panel (c), the energy ranges which can be reached by the {\it Fermi}-GBM (pink), {\it Fermi}-LAT (blue), and Cerenkov Telescope Array (CTA) (green) instruments are illustrated as colored bands. Collisions in which photons with energies above $10^6 \, \mathrm{GeV}$ are able to escape are marked as upward-pointing arrows.}
\end{figure*}

\subsection{Simulation results}

The light curve of the simulated burst is shown in \figu{Figure-2-Bustamante}a, as a black curve.  
Although we show the light curves for only two representative simulations in this study (the aforementioned one and another one with $N_\text{sh} = 100$ and ${\delta t}_{\rm eng} = 0.1$ s, in
\figu{Figure-2-Bustamante}b), we will present a more detailed parameter space study in a future work~\cite{BBMWprep}.
We do not investigate effects of the spectral evolution during the dynamical time for one collision~\cite{Asano:2014nba}, as we imply that taking into account contributions from multiple shells is more relevant, like in the case of gamma rays~\cite{Aoi:2009ty}. 
Note that, although we do not calculate hadronic cascades, their feedback on neutrino spectra is unimportant given the value of the baryonic loading factor used in this work.  

We show in \figu{Figure-3-Bustamante} the neutrino fluence (a), maximal proton energy (b), and maximal gamma-ray escape energy (c) for each collision (dot) as a function of $R_{\text{C}}$.  
The maximal proton energies are obtained from comparing acceleration, dynamical, synchrotron loss, and photohadronic (for protons) timescales.
As a result, we find that the collisions are spread between about $10^6\, \mathrm{km}$ and our choice of $5.5 \cdot 10^{11} \, \mathrm{km}$ for the deceleration radius \cite{Rees:1992ek}, where outflows terminate by the external shock into the circumburst medium. Most collisions occur around $10^{10} \, \mathrm{km}$ --- slightly above the estimate from the geometry \equ{rc}, $R_{\text{C}} \approx 1.6 \cdot 10^9\, \mathrm{km}$. Red dots mark collisions in the neutron-escape model regime (optically thick to $p\gamma$ interactions) and blue empty circles, collisions in the direct proton escape regime. 

Black squares mark subphotospheric collisions, \ie, those for which the Thomson optical depth is larger than unity. The optical depth is obtained by calculating the proton number density from the masses of the shells and assuming that the electron number density is as high as the proton density, which is expected for an electrically neutral plasma. In reality, however, the electron and positron densities may be somewhat higher if there is a significant non-thermal contribution from electron-positron pair production.  
The obtained photospheric radius $R_{\mathrm{ph}} \approx 2 \cdot 10^8 \, \mathrm{km}$ is somewhat larger than the conventional expectation calculated using the dissipated energy in gamma rays ($R_{\mathrm{ph}} \approx 3 \cdot 10^{7} \, \mathrm{km}$).  
This estimate is affected by the efficiency of the conversion from kinetic energy into dissipated energy, which is roughly 25\% in our cases.  However, the more important reason is the significant baryonic loading: since most of the energy is dissipated into protons, the masses of the shells have to be upscaled to match the required energy output in gamma rays ($10^{53} \, \mathrm{erg}$), which leads to larger radii of the photosphere because of higher electron densities. It can therefore be expected that the large baryonic loadings that are needed to describe the UHECR observations~\cite{Murase:2008mr,Baerwald:2014zga} will lead to larger fractions of subphotospheric collisions.

We find that the obtained range of collision radii is large, from under the photosphere out into the deceleration radius, since dissipation occurs for a wide range of $R_{\text{C}}$ especially when the spread of the Lorentz factor $A_\Gamma$ is large~\cite{Beloborodov:2000nn,Li:2008ub}.  
Note that $\lesssim 12$\% of collisions occurs under the photosphere for the chosen parameter set, altogether 118 out of the total 990 collisions, but most of the energy dissipation occurs at large radii $> 10^{10} \, \mathrm{km}$. In the internal shock model, gamma-ray emission should be produced beyond the photosphere, so we only consider collisions beyond the photosphere in the following, unless noted otherwise.
This is conservative, since the baryonic loading may be smaller under the photosphere~\cite{Murase:2008sp} and particle acceleration becomes inefficient for radiation-mediated shocks~\cite{Murase:2013ffa}.
The ratio of total energy emitted as neutrinos via optically-thin internal shocks to the total energy emitted by these collisions as gamma rays is 
4.8\% for this representative parameter set.

Most importantly, \figu{Figure-3-Bustamante}a demonstrates that neutrinos are dominantly produced at small collision radii $R_{\text{C}} \lesssim 10^9 \, \mathrm{km}$, close to the photospheric radius $R_{\rm ph} \approx 2 \cdot 10^{8} \, \mathrm{km}$.  This result can be understood as follows. 
In each collision, the emitted gamma-ray energy, $E_{\gamma-{\rm sh}}^{\rm iso}$, is a fraction of the total dissipated energy. 
The pion production efficiency (fraction of proton energy going into produced pions) at the photon spectral break $\epsilon_{\gamma,\text{break}}^\prime$, which is neglecting spectral effects, can be approximated as~\cite{Murase:2005hy,Murase:2008sp}
\begin{equation}
f_{p\gamma} \propto \frac{\kappa_p \, \sigma_{p \gamma} \, E_{\gamma-{\rm sh}}^{\rm iso}}{4 \pi R_{\text{C}}^2 \Gamma_{\text{m}} \epsilon_{\gamma,\text{break}}^\prime }\,  \, .
\label{equ:fpgamma}
\end{equation}
Here $\Gamma_{\text{m}}$ is the Lorentz factor of the merged shell, $\sigma_{p \gamma}$ is the photohadronic cross section, and $\kappa_p \simeq 0.2$ is the fraction of proton energy going into the pion per interaction. 
Since the internal shock model predicts~\cite{Beloborodov:2000nn} $E_{\gamma-{\rm sh}}^{\rm iso}\propto R_{\text{C}}^{-q}$ for $0\lesssim q \lesssim 2/3$, we expect $f_{p\gamma}\propto R_{\text{C}}^{-2-q}{\epsilon_{\gamma,\text{break}}^\prime}^{-1}$. Hence, since $A_\Gamma$ has to be sufficiently large for efficient energy dissipation ($A_\Gamma = 1$ in the simulations in the present study), neutrino production is typically dominated by collisions at radii around the photosphere. 
UHECR protons come from collisions in the range $10^{8.5} \, \mathrm{km} \lesssim R_{\text{C}} \lesssim 10^{10} \, \mathrm{km}$; see \figu{Figure-3-Bustamante}b. First the maximal proton energy increases with collision radius as (close to the peak) synchrotron losses limit the maximal proton energy, and the magnetic field drops with $R_{\text{C}}$. The peak occurs where adiabatic losses take over, and the decline comes from a decrease of the acceleration timescale for dropping magnetic fields; the expressions for the different energy-gain and energy-loss timescales can be found, \eg, in \Refs~\cite{Murase:2005hy,Baerwald:2013pu}. Note that the UHECRs come from two different components dominating at different collision radii: for $R_{\text{C}} \lesssim 10^{8.5} \, \mathrm{km}$, neutron escape dominates, and for $R_{\text{C}} \gtrsim 10^{8.5} \, \mathrm{km}$, protons directly escaping from the source dominate \--- which are obviously not related to strong neutrino production; see \figu{Figure-3-Bustamante}a. In the chosen example, the main contribution to cosmic rays actually comes from direct escape. Finally, \figu{Figure-3-Bustamante}c illustrates that high 
gamma-ray energies, which can only be observed in CTA or other next-generation imaging atmospheric Cherenkov telescopes, come from large collision radii $R_{\text{C}} \gtrsim 10^9 \, \mathrm{km}$, since for lower radii the optical depth for $\gamma\gamma$ interactions is too high.
As a consequence, neutrinos, cosmic rays, and {\it Fermi}-LAT/CTA gamma rays probe different emission radii. Neutrinos are useful to probe dissipation at small radii, including subphotospheric dissipation.  For dissipation at large radii, where heavy nuclei survive, the TeV gamma-ray diagnostics of a GRB would be useful~\cite{Murase:2008mr}.

\begin{figure}[t]
\begin{center}
\includegraphics[width=1.0\columnwidth,clip=true,trim=0 0 0 0.45cm]{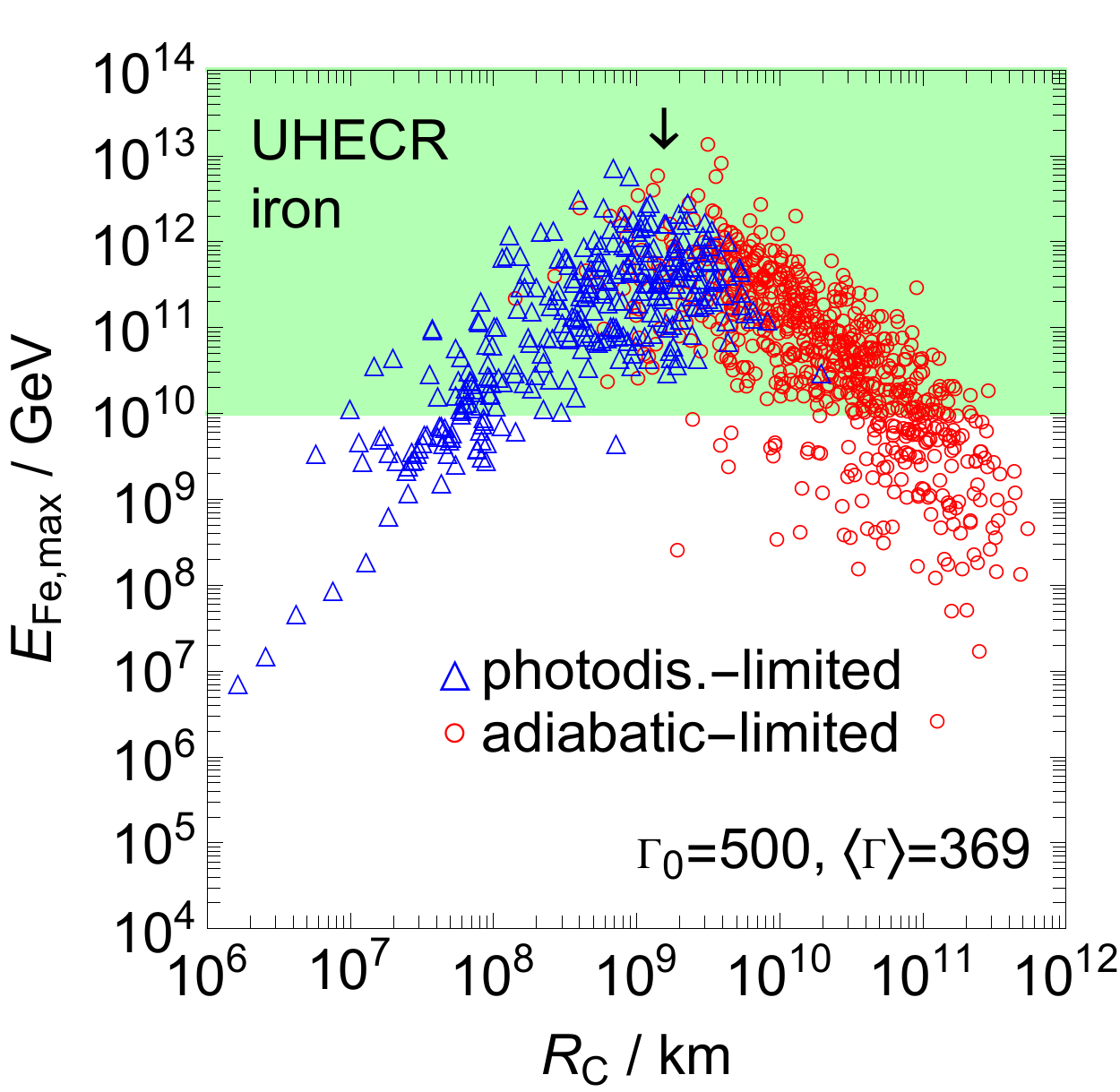} 
\end{center}
\caption{\label{fig:Figure-4-Bustamante}
{\bf Maximum energy to which iron nuclei ($A = 56$, $Z=26$) can be accelerated in each collision of a simulated GRB.} Our standard parameter set is assumed for the simulation. Energy is shown in the source frame, and is calculated for ideal ($\eta=1$) acceleration. Triangles (blue) and circles (red) represent collisions where the energy is limited by break-up due to photodisintegration and by adiabatic losses, respectively. The photodisintegration timescale has been computed using the approximation in \Ref~\cite{Murase:2008mr}. The ultra-high energy range for cosmic rays, above $10^{10}$ GeV, is shown as a green band. The downward-pointing arrow marks the approximate energy above which adiabatic energy losses dominate.}
\end{figure}

There has been some evidence that the composition of UHECRs is heavy~\cite{Abraham:2010yv}. Initial studies such as \Refs~\cite{Lemoine:2002vg,Pruet:2002hi,Beloborodov:2002af} concluded that heavy nuclei cannot survive inside GRBs: photodisintegration on fireball photons would break them up into lighter nuclei and protons. \Ref~\cite{Anchordoqui:2007tn} calculated the neutrino emission from the injection of both protons and nuclei and found that the latter cannot survive in internal shocks; however, only collisions at very small radii, around $10^8$ km, were considered. It has been argued that the typical collision radius is much larger (see, \eg, \Ref~\cite{Zhang:2010jt} and references therein) and that heavy nuclei can be largely loaded in GRB jets~\cite{Metzger:2011xs,Horiuchi:2012by}. Therefore, acceleration of nuclei to ultra-high energies and their survival against photodisintegration are possible, provided $R_{\text{C}}$ is large enough~\cite{Murase:2008mr,Wang:2007xj,Globus:2014fka}. In \figu{Figure-4-Bustamante} we show that this is indeed the case for our simulations. The figure shows the maximum energy to which iron nuclei ($A = 56$, $Z=26$) can be accelerated at each of the collisions. The energy is a factor of $26$ higher than for protons (compare to \figu{Figure-3-Bustamante}b), where its absolute magnitude is a consequence of the assumed acceleration efficiency. Here the photodisintegration timescale has been calculated using the approximation in \Ref~\cite{Murase:2008mr}. Triangles (blue) and circles (red) represent collisions in which the maximum energy is limited, respectively, by the break-up of the nucleus due to photodisintegration and by adiabatic losses. Even though photodisintegration losses dominate up to $\sim 10^{9}$ km, after which adiabatic losses take over, maximum energies well within the UHE band can be achieved at the turning point, where most of the UHECR emission would come from. Note that this turning point is about a factor of five higher in $R_{\text{C}}$ than for protons (compare arrows in \figu{Figure-3-Bustamante}b and \figu{Figure-4-Bustamante}), which means that UHECR nuclei on average reach their peak energy at higher $R_{\text{C}}$ than UHECR protons. UHECR nuclei may also escape directly at the highest energies, but there is no such thing as neutron escape. It is therefore expected that nuclei come from somewhat larger collision radii than protons at the highest energies, where the radiation densities are too low to break up the nuclei. Since the actual energy output of heavy nuclei depends on the nuclear loading (\ie, an additional assumption is required), we do not show their energy output explicitly in the following.

\begin{figure*}[t]
\begin{center}
\includegraphics[width=\textwidth,clip=true,trim=0 1.5cm 0 1.5cm]{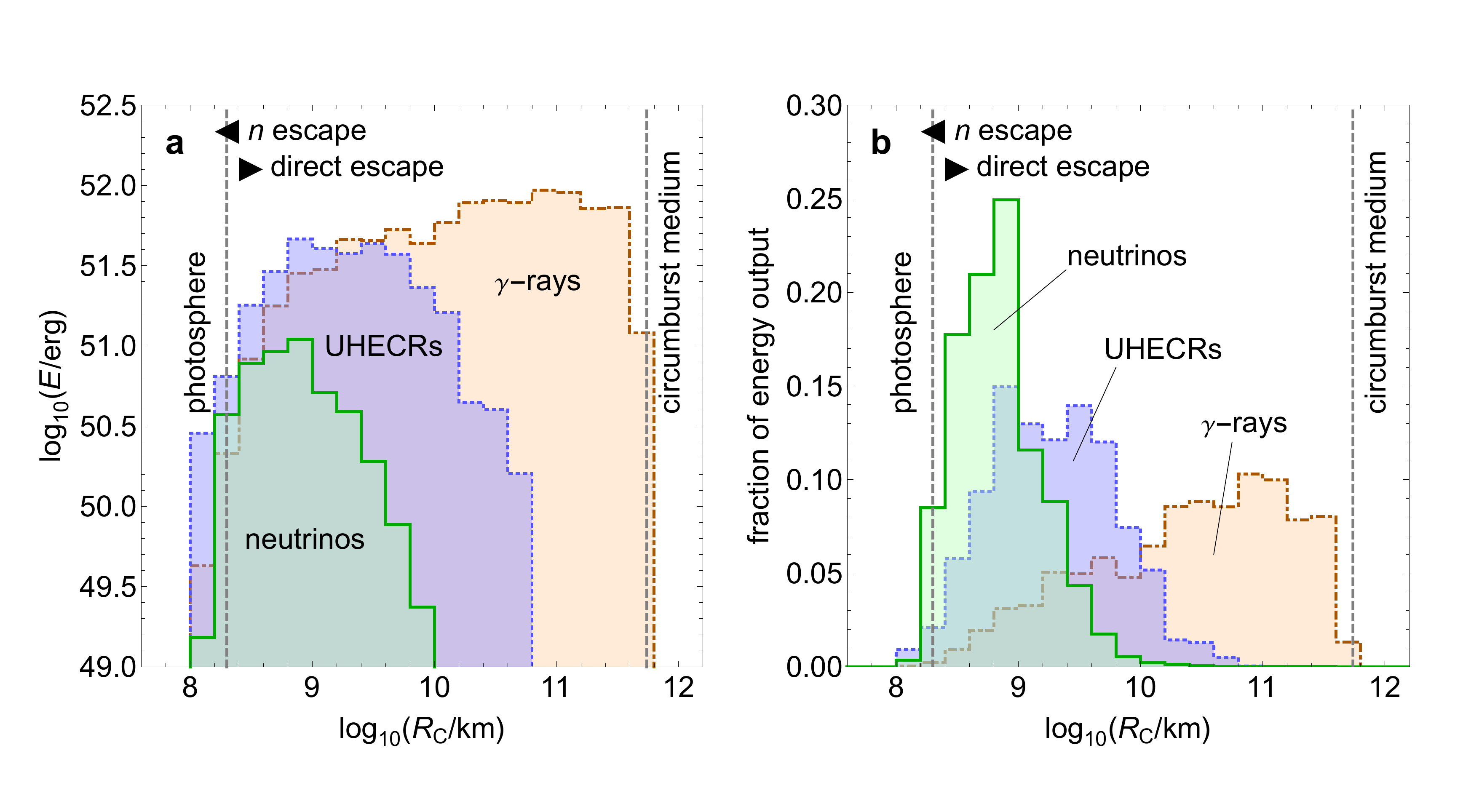}
\end{center}
\caption{\label{fig:Figure-5-Bustamante}{\bf Energy dissipated in (prompt) gamma rays, neutrinos (all flavors), and CR protons (UHECRs from $10^{10}$ to $10^{12} \, \mathrm{GeV}$) beyond the photosphere.} Energies are binned as a function of the collision radius. Panel (a) shows absolute energy values; panel (b) shows the fraction of energy output normalized to one for each messenger. Neutron escape dominates the cosmic ray emission below $R_{\text{C}} \approx 10^{8.5} \, \mathrm{km}$, while proton escape dominates above this radius. The rough value of the photospheric radius and the assumed radius of the circumburst medium are indicated as dashed lines.}
\end{figure*}

In order to obtain an even more quantitative statement of how much energy is released as a function of collision radius, we show in \figu{Figure-5-Bustamante} binned distributions for the prompt gamma rays, neutrinos, and cosmic-ray protons, which are all directly calculable within our model. \figu{Figure-5-Bustamante}a shows the energy output per bin, while \figu{Figure-5-Bustamante}b shows the fraction of energy in each bin compared to the total, for each particle species. We note that the energy per messenger per bin is obtained as a product of energy released per collision, and the number of collisions occurring per $R_{\text{C}}$-bin; especially the latter number is important to get the proper weighing of $R_{\text{C}}$.  The result confirms the above observations: the neutrino production is dominated by small values of $R_{\text{C}}$ just beyond the photosphere from within a relatively narrow region $R_{\text{C}} \approx 10^{8.5}$ to $10^{9} \, \mathrm{km}$, the cosmic-ray production by intermediate $R_{\text{C}} \approx 10^9$ to $10^{10} \, \mathrm{km}$, and the prompt gamma-ray emission is, in fact, dominated by large $R_{\text{C}}$, at around $10^{10}$ to $10^{11.5} \, \mathrm{km}$ -- compatible with what is typically expected in the literature~\cite{Nakar:2002gd}.  
These results have significant implications: our knowledge of the prompt phase of GRBs is obtained from gamma rays, of course, and, consequently, $R_{\text{C}}$ is derived from gamma-ray observations. This collision radius is, however, not the one to be used for neutrino or cosmic-ray calculations. It is therefore conceivable 
that multi-zone predictions are different from the naive one-zone expectation based on the gamma-ray emission radius.  One can also read off from \figu{Figure-5-Bustamante} that a significant amount of energy in UHECRs is transported away by direct escape, unrelated to neutrino production, which may affect the predicted neutrino flux if normalized to the observed UHECRs, as in, \eg, \Ref~\cite{Baerwald:2014zga}.

\begin{figure*}[t]
\begin{center}
\includegraphics[width=\textwidth]{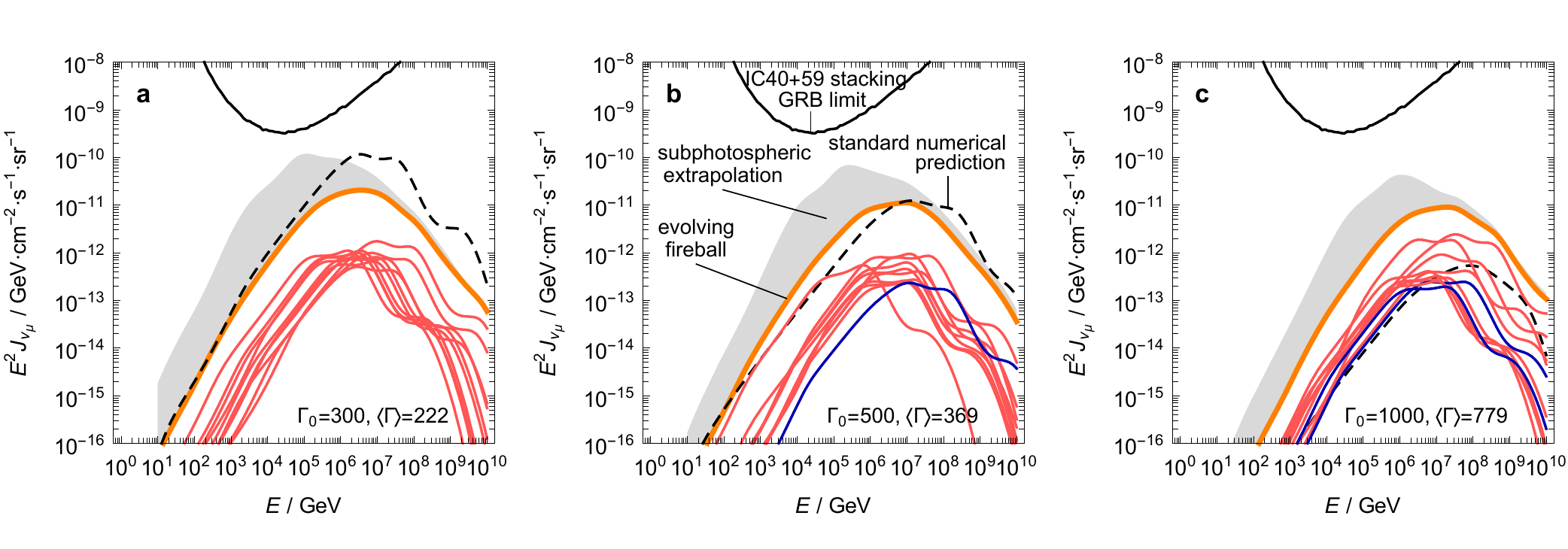}
\end{center}
\caption{\label{fig:Figure-6-Bustamante}{\bf Quasi-diffuse neutrino spectra from simulations of multiple internal shocks.} Muon-neutrino spectra ($\nu_\mu+\bar{\nu}_\mu$) from collisions beyond the photosphere (thick orange curves), reference spectra computed from averaged burst parameters in the conventional approach (dashed curves), and maximal subphotospheric extrapolations (shaded regions) for three different values of $\Gamma_0$ in the different panels: 300 (a), 500 (b), and 1000 (c). The individual (dominant) collisions (contributing to the thick orange curves) are shown also as thin red and blue curves corresponding, respectively, to the optically thick to $p\gamma$ interactions regime, with the neutron escape dominating at the maximum energy, and to the regime dominated by direct proton escape instead. The (thick orange) spectra (``evolving fireball'') are estimates of the diffuse flux obtained from the single-burst fluence $\mathcal{F}$ (one GRB at $z=2$) by assuming $\dot N=667 \, \mathrm{yr}^{-1}$ long bursts per year over the whole sky ($J=\mathcal{F} \times \dot N \times (4 \pi)^{-1}$). The diffuse GRB flux limit from the IC40+59 analysis~\cite{Abbasi:2012zw} is shown as a thin black curve. The obtained average values $\langle \Gamma \rangle$ from the simulation, corresponding to the observable $\Gamma$, are depicted as well.}
\end{figure*}

We show in \figu{Figure-6-Bustamante} the predicted quasi-diffuse neutrino spectra from collisions beyond the photosphere as thick orange curves for three different values of $\Gamma_0$, where \figu{Figure-6-Bustamante}b corresponds to our standard assumptions.  Note that the neutrino fluence per burst has been rescaled to a quasi-diffuse flux prediction by assuming 667 (identical) bursts per year and is significantly below the current diffuse neutrino signal reported by IceCube at the level of $10^{-8} \, \text{GeV} \, \text{cm}^{-2} \, \text{s}^{-1} \, \text{sr}^{-1}$ flux~\cite{Aartsen:2014gkd}.  The dashed curves correspond to the standard assumption that all collisions occur at the same radius, derived from gamma-ray observations.  To generate these curves, we use the parameters  $N_\text{coll}$, $t_{\text{v}}$,  $\langle\Gamma\rangle$, and $T$ obtained from the simulation assuming identical shells with a collision radius obtained from \equ{rc} ($R_{\text{C}} \approx 10^{9.2} \, \mathrm{km}$ in \figu{Figure-6-Bustamante}b).  The reference flux in \figu{Figure-6-Bustamante}b is significantly lower than the prediction in \Ref~\cite{Hummer:2011ms}. In that reference, the same parameters as in the IceCube analysis~\cite{Abbasi:2011qc} were used for comparison, implying that $R_{\text{C}} \approx 1.9 \cdot 10^8 \, \mathrm{km}$. That is about one order of magnitude smaller than the $R_{\text{C}}$ used here; \cf, \equ{fpgamma} for its impact on the neutrino flux. The reference flux in \figu{Figure-6-Bustamante}a is comparable to \Ref~\cite{Hummer:2011ms}.

We first of all find that the neutrino spectra from collisions beyond the photosphere (thick orange curves) all exhibit the same flux level quite independently of $\Gamma_0$ (and even of $A_\Gamma$, as we have explicitly tested). The expected neutrino flux per flavor is at the level of $E^2 J \sim 10^{-11} \, \mathrm{GeV \, cm^{-2} \, sr^{-1} \, s^{-1}}$, peaking between $10^5$  and $10^7 \, \mathrm{GeV}$.  This contribution can be regarded as a minimal prediction for the neutrino flux, as it can be inferred from gamma-ray observations and hardly depends on the parameters.  Note that this flux is probably outside the  sensitivity of the existing IceCube experiment, but it will provide a target for the optimization of the planned high-energy volume upgrade.
There is a significant qualitative difference to conventional models such as \Refs~\cite{Waxman:1997ti,Guetta:2003wi}, for which the pion production efficiency contains a factor $\Gamma^{-4}$ coming from the collision radius estimate in \equ{rc} applied to \equ{fpgamma}. However, the optical thicknesses to Thomson scattering and photohadronic interactions both scale $\propto R_{\text{C}}^{-2}$, which leads to the following estimate for the pion production efficiency {\it at the photosphere} independent of $\Gamma$~\cite{Murase:2008sp}:
\begin{equation}
f_{p \gamma}^{\mathrm{ph}} \sim 5 \times \frac{\varepsilon}{0.25} \times \frac{\epsilon_e}{0.1} \times \frac{1 \, \mathrm{keV}}{\epsilon_{\gamma,\text{break}}^\prime} \,.
\label{equ:pgph}
\end{equation}
Here $\epsilon_e$ is the fraction of the dissipated energy going into photons, and $\varepsilon$ is the dissipation efficiency (ratio between dissipated and kinetic energies). Notably, $\Gamma$ drops out of
$f_{p \gamma}^{\mathrm{ph}}$  -- unless the break energy is fixed in the observer's frame, in which case there is a single factor of $\Gamma$ retained.

When the innermost collisions give the dominant contributions, the time-integrated neutrino fluence roughly scales as
\begin{equation}
\mathcal{F}_\nu \propto \frac{N_{\rm coll}(f_{p\gamma} \gtrsim 1)}{N_{\rm coll}} \times {\rm min}[1,f_{p \gamma}^{\mathrm{ph}}] \times \frac{\epsilon_p}{\epsilon_e} \times  E_{\mathrm{iso}},
\label{equ:fnu}
\end{equation}
where $N_{\rm coll}(f_{p\gamma} \gtrsim 1)$  is the number of collisions with efficient neutrino production close to the photosphere, $N_{\rm coll}$ is the total number of collisions, and $\epsilon_p$ is the fraction of energy going into protons.
Since the number of dominant collisions contributing to the neutrino flux is of order ten almost independently of the model parameters (see thin solid curves in \figu{Figure-6-Bustamante}), the neutrino flux prediction is relatively robust.
The neutrino prediction above the photosphere hardly depends on the baryonic loading ($\epsilon_p/\epsilon_e$) as well, as long as most of the energy is dissipated into protons. Increasing the baryonic loading in \equ{fnu} is compensated by a correspondingly smaller $\epsilon_e$ in \equ{pgph}. As a result, the neutrino flux is roughly independent of $\epsilon_p/\epsilon_e$  -- which we have explicitly tested numerically. 

\begin{figure*}[t]
\begin{center}
\includegraphics[width=\textwidth]{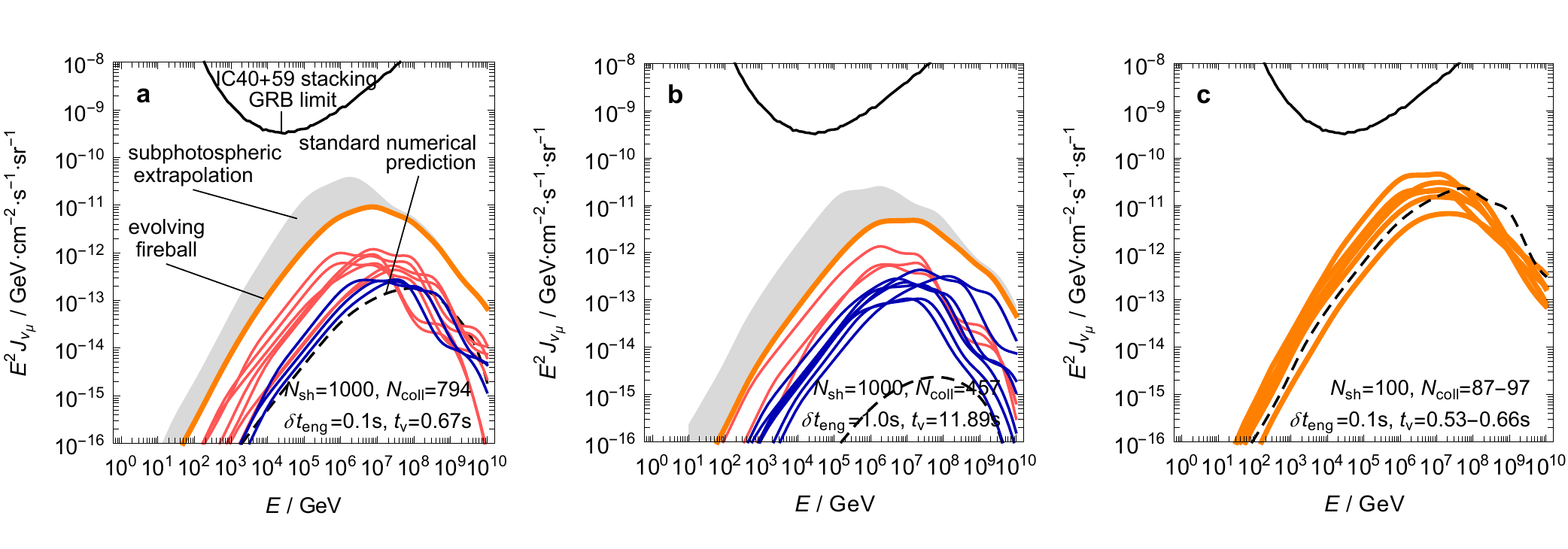}
\end{center}
\caption{\label{fig:Figure-7-Bustamante}
{\bf Quasi-diffuse muon-neutrino spectra from simulations with alternative parameter sets.} The $\nu_\mu+\bar{\nu}_\mu$ neutrino spectra in this figure should be compared to the ones obtained using our standard parameter set, \figu{Figure-6-Bustamante}b (${\delta t}_{\rm eng}=0.01 \, \mathrm{s}$, $N_\mathrm{sh}=1000$, $\Gamma_0=500$). Panels (a) and (b) show two larger values of ${\delta t}_{\rm eng}$ for $N_\mathrm{sh}=1000$, which leads to a longer burst duration $T$. Panel (c) uses ${\delta t}_{\rm eng}=0.1 \, \mathrm{s}$ with a reduced number of shells $N_\mathrm{sh}=100$ (corresponding to the light curves in \figu{Figure-2-Bustamante}b), \ie, $T$ is similar to the original example but the light curve is less spiky because of fewer collisions. Since in that case the statistical fluctuations from burst to burst increase, we show six different realizations of the predicted flux.}
\end{figure*}

We have also tested that this prediction does not depend on the variability timescale of the burst: \figu{Figure-7-Bustamante} shows predictions for two different values of the emitter uptime ${\delta t}_{\rm eng} = 0.1$ s (a) and $1$ s (b), where the fixed $N_\mathrm{sh}=1000$ leads  to a longer burst duration $T$. Clearly, the quasi-diffuse flux coming from the simulations is independent of the value of $t_{\text{v}}$, as expected from \equ{pgph}. This markedly contrasts with the standard numerical prediction, in which larger variability timescales unavoidably imply lower particle densities at the source and, therefore, a reduced neutrino production. In \figu{Figure-7-Bustamante}c, ${\delta t}_{\rm eng}=0.1 \, \mathrm{s}$ with a reduced number of shells $N_\mathrm{sh}=100$ is chosen, corresponding to the light curve in \figu{Figure-2-Bustamante}b, that has a similar duration as the light curve for our standard simulation, but fewer pulses. In this case, the obtained result depends on the actual realization of the $\Gamma$-distribution, as only a few collisions dominate the neutrino flux and lead to strong variations \--- six different realizations of the $\Gamma$-distribution are shown in \figu{Figure-7-Bustamante}c. As expected, our conclusions hold for a sufficiently large sample of GRBs centered around our average prediction.
We expect that these fluctuations become more severe for even fewer pulses from fewer collisions, as it has been studied for the contributions from different bursts in \Sec~4 of \Ref~\cite{Baerwald:2011ee}.
The independence on the model parameters implies that the predicted flux $E^2 J \sim 10^{-11} \, \mathrm{GeV \, cm^{-2} \, sr^{-1} \, s^{-1}}$ is very robust. The only exception may be increasing $E_{\mathrm{iso}}$ (see \equ{fnu}) or the baryonic loading, which may in fact be required to match the injected energy needed to describe UHECR observations; see \Sec~2 in \Ref~\cite{Baerwald:2014zga}.

The photon spectra can still be approximated by the Band function up to a Thomson optical thickness of ten or so~\cite{Pe'er:2008te,Beloborodov:2010zb}, which occurs under the photosphere. This means that we can extrapolate our assumptions to below the photosphere to some degree. In the most extreme case, all energy may be dissipated into neutrinos, whereas the energy of neutrons, protons, and gamma rays is reconverted into kinetic energy -- this is, however, very speculative, as nonthermal particle acceleration may not occur efficiently~\cite{Murase:2008sp}. We show the corresponding subphotospheric extrapolations for the neutrino spectra as highly uncertain shaded regions in \figu{Figure-6-Bustamante}, corresponding to the contribution to the black squares in \figu{Figure-3-Bustamante}. Since the photospheric radius increases with decreasing $\Gamma$, the number of subphotospheric collisions increases with it, and their contribution in \figu{Figure-6-Bustamante}a can be much higher than in \figu{Figure-6-Bustamante}b (and in \figu{Figure-6-Bustamante}c much lower). As a consequence, the subphotospheric extrapolation may even reach the current sensitivity limit, and can be already constrained with current data. However, note again that this extrapolation is highly uncertain, as gamma-ray data cannot be used to obtain information about below the photosphere.

Finally, we show the ``neutrino light curve'' for our standard parameter set in \figu{Figure-2-Bustamante}a as a dotted (red) curve; \figu{Figure-2-Bustamante}b shows it for a simulation with fewer collisions and longer emitter uptime, corresponding to the neutrino spectra in \figu{Figure-7-Bustamante}c.  It can be clearly seen that the neutrino flux is typically much lower than the gamma-ray flux except in some rare cases where the collision occurs close to the photosphere. Furthermore, the variation of the neutrino flux is larger due to the strong dependence of the pion production efficiency on $R_{\text{C}}$. 
One qualitative prediction that could help neutrino searches is that neutrinos are more likely to be associated with gamma-ray spikes that are pulses with very short variability timescales.

\section{Discussion}

In summary, we have studied neutrino, gamma-ray (at different energies), and cosmic-ray production in an evolving GRB outflow.  We have demonstrated that they are produced at different collision radii.  Consequently, the typical emission radius derived from prompt gamma rays cannot be directly applied to neutrino and UHECR production, and the GRB will look very different from the point of view of different messengers.  This concept is well known from conventional astronomical observations, where astrophysical objects look very different in different wavelength bands.

The neutrino spectra derived from gamma-ray observations are dominated by the emission close to the photosphere  at $R_{\text{C}} \approx 10^{8.5}$ to $10^{9} \, \mathrm{km}$, as the pion production efficiency depends on the collision radius in a non-linear way. 
UHECR protons have been shown to be produced at intermediate collision radii $R_{\text{C}} \approx 10^{8.5}$ to $10^{10} \, \mathrm{km}$, where the magnetic fields are high enough for efficient acceleration, but not so high that synchrotron losses limit the maximal proton energies. We have taken into account two possibilities for UHECR escape: emission as neutrons, which are not magnetically confined, and emission as protons from the edges of the shells -- the dominant mechanism in each collision depends on the parameters of the colliding shells. Since the neutrons come from photohadronic interactions, their production dominates at smaller collision radii, where the $p\gamma$ optical depth is higher, whereas protons tend to be directly emitted at large radii. Heavier nuclei can also survive for sufficiently large collision radii; their actual contribution depends on the nuclear loading.
The main energy in gamma rays is deposited between around $10^{10}$ to $10^{11.5} \, \mathrm{km}$, compatible with earlier estimates. In particular, gamma rays at the highest energies, such as in the energy range only accessible to CTA, cannot come from collision radii $\lesssim 10^{9} \, \mathrm{km}$ as the photon 
densities are too high  there to let them escape. 

For the quasi-diffuse neutrino flux prediction, we have identified two distinctive contributions. Above the photosphere, gamma-ray observations can be used to infer the pion production efficiency, which leads to a neutrino flux per flavor $E^2 J \sim 10^{-11} \, \mathrm{GeV \, cm^{-2} \, sr^{-1} \, s^{-1}}$ for an assumed isotropic energy of $10^{53} \, \mathrm{erg}$ emitted in gamma rays.  Especially, there is no strong dependence on the Lorentz boost $\Gamma$, in contrast to conventional one-zone models, as both the photosphere and the pion production efficiency scale with the collision radius in the same way. This is the minimal neutrino flux which one would expect in stacking analyses based on the actual gamma-ray observations, such as \Ref~\cite{Abbasi:2012zw}. There is also a significantly milder dependence on the baryonic loading, as this parameter changes the photosphere of the model at the same time that it rescales the neutrino flux. The prediction hardly depends on the time variability or number of pulses in the GRB light curve within a certain time window either. However, if the overall number of pulses is low, these will only come from a very small number of collisions, which means that large statistical fluctuations of the neutrino flux from burst to burst are expected even for the same parameter values. In that case, our observations have to be instead interpreted for a large enough ensemble of bursts. Note that the chosen isotropic energy and baryonic loading may not be sufficient to describe UHECR observations, see \Sec~2 of \Ref~\cite{Baerwald:2014zga} for a detailed discussion, which will need to be addressed in a future study.

The neutrino flux is significantly lower than earlier predictions~\cite{Hummer:2011ms} because a) we have explicitly excluded subphotospheric contributions, b) large photospheric radii have been obtained as a consequence of significant baryonic loadings (ten) and the moderate energy dissipation efficiency of the fireball (25\%), and c) only a small number of collisions beyond the photosphere occurs at radii where the neutrino production efficiency is high.  This expected ``minimal'' flux is beyond the sensitivity of the current IceCube experiment, but could be reached in future high-energy extensions~\cite{Aartsen:2014njl}.  No gamma-ray information from deep below the photosphere can be directly obtained, and the neutrino production in that regime is more speculative~\cite{Murase:2013ffa}. In principle, however, a high-energy extension of the detector could also constrain the subphotospheric neutrino production.

Our results imply that model-dependent studies of the multi-messenger connection, such as a GRB stacking analysis of neutrino fluences, can be improved, and give a stronger case for testing the hypothesis that UHECRs originate from GRBs.  Compared to the one-zone model, some additional assumptions need to be made for the distribution of the collision radii.  In particular, the width of the initial distribution of the bulk Lorentz factor $A_\Gamma$, with which the shells are set out by the central engine, turns out to be the key additional parameter.  It can in principle be obtained from comparing the light curves between simulation and observation.  On the other hand, we have the advantage that the uncertainty in $R_{\text{C}}$, which is the key issue in the standard model, disappears, as a collision radius distribution is now predicted by the theory.  While we expect that the bulk Lorentz factor distribution has to be broad in some way to maintain a high dissipation efficiency, it remains to be studied how the results change for qualitatively different distributions. There should also be new opportunities stemming from our results: different messengers can be used to study different regions of an evolving GRB outflow. For instance, direct neutrino and gamma-ray observations, in CTA, of a single GRB would open windows to very different regions of the GRB.


During completion of this work, \Ref~\cite{Globus:2014fka} appeared, which shares some common aspects.

\vspace*{0.2cm}

{\bf Acknowledgements.}~We thank V.~Mangano, E.~Waxman, and B.~Zhang for discussion and comments.  
This work is supported by NASA through Hubble Fellowship Grant No.~51310.01 awarded by the STScI, which is operated by the Association of Universities for Research in Astronomy, Inc., for NASA, under Contract No.~NAS 5-26555 (K.M.). M.B.~and W.W.~would also like to acknowledge support from DFG grants WI 2639/3-1 and WI 2639/4-1, and the ``Helmholtz Alliance for Astroparticle Physics HAP'',  funded by the Initiative and Networking Fund of the Helmholtz Association. P.B.~acknowledges support from NASA grant NNX13AH50G. M.B., K.M., and W.W.~would like to thank the Kavli Institute for Theoretical Physics at UCSB for its hospitality during the development of part of this work. This research was supported in part by the National Science Foundation under Grant No.~NSF PHY11-25915.

\vspace*{0.2cm}

{\bf Author contributions.}~All authors contributed to all aspects of this work, discussed the results, and commented on the manuscript.

\vspace*{0.2cm}

{\bf Competing financial interests.}~The authors declare no competing financial interests.


%

\end{document}